\documentclass[prl,superscriptaddress,twocolumn,showpacs]{revtex4}

\usepackage{amsfonts,amssymb,amsmath}
\usepackage[]{graphics,graphicx,epsfig}
\usepackage{amsthm}

\def\identity{\leavevmode\hbox{\small1\kern-3.8pt\normalsize1}}

\renewcommand{\epsilon}{\varepsilon}

\begin{document}

%\preprint{APS/123-QED}

\title{Particle addition and subtraction as a test of bosonic quality}

\author{Pawe\l\ Kurzy\'nski}
\affiliation{Centre for Quantum Technologies,
National University of Singapore, 3 Science Drive 2, 117543 Singapore,
Singapore}
\affiliation{Faculty of Physics, Adam Mickiewicz University,
Umultowska 85, 61-614 Pozna\'{n}, Poland}

\author{Ravishankar Ramanathan}
\affiliation{Centre for Quantum Technologies,
National University of Singapore, 3 Science Drive 2, 117543 Singapore,
Singapore}

\author{Akihito Soeda}
\affiliation{Centre for Quantum Technologies,
National University of Singapore, 3 Science Drive 2, 117543 Singapore,
Singapore}

\author{Tan Kok Chuan}
\affiliation{Centre for Quantum Technologies,
National University of Singapore, 3 Science Drive 2, 117543 Singapore,
Singapore}

\author{Dagomir Kaszlikowski}
\email{phykd@nus.edu.sg}
\affiliation{Centre for Quantum Technologies,
National University of Singapore, 3 Science Drive 2, 117543 Singapore,
Singapore}
\affiliation{Department of Physics,
National University of Singapore, 3 Science Drive 2, 117543 Singapore,
Singapore}

\date{\today}% It is always \today, today,
             %  but any date may be explicitly specified

\begin{abstract}

We propose a test to measure the bosonic quality of particles with respect to physical operations of single-particle addition and subtraction. We apply our test to investigate bosonic properties of composite particles made of an even number of fermions and suggest its experimental implementation. Furthermore, we discuss the features of the processes of particle addition and subtraction in terms of optimal quantum operations. 

\end{abstract}

% PACS, the Physics and Astronomy
                             % Classification Scheme.
%\keywords{Suggested keywords}%Use showkeys class option if keyword
                              %display desired
\maketitle

{\it Introduction.} Non-commutativity of photonic addition and subtraction was  recently confirmed in the laboratory \cite{Parigi,Zavatta}. An initial thermal state of light was subjected to both single photon addition and subtraction, mathematically denoted by bosonic creation and annihilation operators, $a^{\dagger}$ and $a$, respectively. Two scenarios were considered, photonic addition followed by subtraction (AS) and vice versa (SA). According to quantum theory, the final states in the two scenarios should differ due to the bosonic commutation rule $[a,a^{\dagger}]=\openone$. In general
\begin{eqnarray}
& &\rho_{SA}\equiv(a^{\dagger}a)\rho (aa^{\dagger})=\sum_{n,m}\rho_{n,m} n m|n\rangle\langle m|\neq \label{e1} \\ & & \rho_{AS}\equiv(aa^{\dagger})\rho (a^{\dagger}a)=\sum_{n,m}\rho_{n,m} (n+1)(m+1)|n\rangle\langle m|, \nonumber
\end{eqnarray}
where $|n\rangle$ denotes the state of $n$ indistinguishable particles in the same mode; for simplicity the states are not normalized. 

Here, we show that bosonic addition and subtraction can be used not only to verify the commutation relations of ideal bosons, but also to detect and infer some structural properties of composite particles (consisting of an even number of fermions) as well as to test to what extent they behave like ideal bosons. The paper is organized as follows. First, we introduce a simple test to measure bosonic quality based on single particle addition followed by subtraction. Since we are not directly interested in testing commutation relations we simplify the previous scenarios and do not consider the process in which subtraction is followed by addition. This also allows us to bypass one important problem, namely that the operations of addition and subtraction do not commute even for classical distinguishable particles. We then discuss general quantum operations (channels) that efficiently add and subtract a particle. We identify important criteria that have to be satisfied by bosonic addition/subtraction channels and apply these channels to composite particles. We find that our measure of bosonic quality depends on the entanglement between fermions, therefore it provides us with information about the internal structure of the composite particle and can be used as an entanglement witness. We conclude with a discussion of how to implement bosonic addition and subtraction of composite particles in the laboratory.

{\it Measure of bosonic quality.} In this work, we propose to compare the resulting state after particle addition followed by subtraction with the initial state of the particles. Note that the operation AS for distinguishable particles, described by the creation operator $a_d^{\dagger}=\sum_{n=0}^{\infty}|n+1\rangle\langle n|$ and the corresponding annihilation operator, leaves the initial state unchanged unlike in the case of ideal bosons. However, the action of SA changes even the state of distinguishable particles. A successful subtraction indicates the absence of vacuum in the initial state and one has to update and re-normalize the state of the system. This lower bound on the number of particles is the heart of the classical lack of commutation of particle addition and subtraction $[a_d,a_d^{\dagger}]=|0\rangle\langle 0|$. In order to get rid of this classical component in a test of bosonic commutation relations, one has to consider states that do not contain any vacuum. Since both states in \cite{Parigi, Zavatta} were thermal states having a substantial occupancy of vacuum (exceeding $50\%$), it would be interesting to repeat these experiments on vacuum-less states.

Apart from the consideration of just the AS operation and the initial state, we make one further simplification for the sake of experimental convenience. Note that in general, to compare two states (for example using the fidelity measure as done in \cite{Parigi, Zavatta}), it is necessary to perform state tomography which is experimentally an extremely demanding process. The calculation of fidelity is however not necessary to detect the change caused by AS, it is sufficient to measure the probability distribution of the number of particles $\{p_0,p_1,\dots\}$, where $p_n$ denotes the probability of detecting $n$ particles. This is still complicated due to the fact that the current particle detectors do not efficiently measure the number of particles. Instead, their action can be described by measurement operators $\{|0\rangle\langle 0|, \openone - |0\rangle\langle 0|\}$ (see additional material in \cite{Parigi}). These detectors allow the measurement of $p_0$, which is sufficient to detect the change due to AS, and consequently, to provide a measure of bosonic quality. 

The action of bosonic AS affects the probability distribution in the following manner: $p_n \rightarrow (n+1)^2 p_n$, which together with normalization implies a decrease in $p_0$. In contrast, the AS operation for distinguishable particles leaves $p_0$ unchanged. Due to the normalization the change in $p_0$ depends on the total probability distribution $\{p_n\}$. Let us now examine the difference in $p_0$, 
\begin{equation}
p_0-p_0^{AS}=p_0 - \frac{p_0}{\sum_{k=0}^{n_{max}}(k+1)^2 p_k}=p_0-\frac{p_0}{\langle (N+1)^2 \rangle}, \nonumber
\end{equation} 
where $p_0^{AS}$ denotes the vacuum occupancy after AS, $N$ denotes the particle number operator and $n_{max}$ denotes the maximum number of particles in the system. Simple optimization yields that the maximal change occurs for $p_0=\frac{n_{max}+1}{n_{max}+2}$ and $p_{n_{max}} = 1 - p_0$. 
The greater the maximal number of particles $n_{max}$, the greater the change in $p_0$ after AS. At this point, it is clear that the above difference can be taken as a measure of bosonic quality $M=p_0-p_0^{AS}$. For distinguishable particles this is always zero, whereas for true bosons this is $M_b=\frac{n_{max}}{n_{max}+2}$, approaching one in the limit of large $n_{max}$.
We propose to consider the simplest optimal probability distribution taking into account only the vacuum and a single particle component $\{p_0=\frac{2}{3},p_1=\frac{1}{3}\}$, for which $M_b=\frac{1}{3}$. Note that the mixture $p_0|0\rangle\langle 0| + (1-p_0) |1\rangle\langle 1|$ can be prepared for any type of particle (even for fermions). The values $M=0$ and $M=\frac{1}{3}$ correspond to distinguishable particles and bosons, respectively. For simplicity, we define our measure of bosonic quality as 
\begin{equation}\label{measure}
{\cal M}=2-3p_0^{AS}.
\end{equation}
For ideal bosons, ${\cal M}_b=1$ and for distinguishable particles ${\cal M}_d=0$. Any value between $0$ and $1$ corresponds to imperfect bosons.

{\it Addition and subtraction channels.} We now move to some general considerations of the particle addition and subtraction operations. In general, the processes of particle addition and subtraction are not deterministic. Moreover, they cannot be formulated simply as Kraus operators $K_j$ which describe nondeterministic evolutions in terms of completely positive quantum channels $\rho'=\sum_j K_j\rho K_j^{\dagger}$ \cite{NC}. The reason is that all the Kraus operators $\{K_0,K_1,\dots\}$ which describe a quantum channel cannot increase the norm of the state, i.e., $\sum_j K_j^{\dagger}K_j\leq\openone$. Setting $K_0=a^{\dagger}$ yields $K_0^{\dagger}K_0=aa^{\dagger}=N+1$. The eigenvalues of the bosonic particle number operator $N$ lie in the set of all nonnegative integers, which together with the requirement that the norm cannot increase immediately implies the negativity of the remaining operators $K_j^{\dagger}K_j$ for $j\neq 0$. The case of the annihilation process is analogous. It is interesting to note that the probabilistic nature of the operators $a^{\dagger}$ and $a$ can also be deduced from the fact that deterministic addition and subtraction could lead to an increase of entanglement via local operations. This can be seen for example by considering the state of a single particle in two modes $A$ and $B$, $|\psi \rangle$ = $\alpha |0_{A} 1_{B}\rangle + \beta |1_{A} 0_{B}\rangle$ \cite{PKOK} with real parameters $\alpha^{2}$ + $\beta^{2}$ = $1$ and $\alpha^2 >  2 \beta^2$, that has entanglement measured by concurrence given as $2 \alpha \beta$. It is clear that a local operation of addition followed by subtraction at mode $A$ leads to the state $|\psi_{AS} \rangle$ = $\frac{1}{\alpha^2 + 4 \beta^2} (\alpha |0_{A} 1_{B} \rangle + 2 \beta |1_{A} 0_{B} \rangle)$ with entanglement measured by concurrence given as $\frac{4 \alpha \beta}{\alpha^2 + 4 \beta^2}$ which is larger than the initial entanglement. The probabilistic nature of the operators $a^{\dagger}$ and $a$ is needed to ensure that entanglement does not increase via local operations.

Note that any operator which effectively adds one particle to the system is of the form
\begin{equation}\label{eff}
a^{\dagger}_{eff}=\sum_{n=0}^{\infty}f_n|n+1\rangle\langle n|. 
\end{equation}
An effective annihilation operator is the hermitian conjugate of the above. This operator is a valid Kraus operator if $|f_n|^2\leq 1$ for all $n$. It is convenient to rewrite it in the following form
\begin{equation}
a^{\dagger}_{eff}=g(N)a^{\dagger}=\sum_{n=0}^{\infty}g(n+1)\sqrt{n+1}|n+1\rangle\langle n|, \nonumber
\end{equation}
where $g(N)$ is a function of the particle number operator. The extreme case in which all multiplicative factors are equal to one corresponds to the creation operator of distinguishable particles $a^{\dagger}_{d}$,
where $g(N)=1/\sqrt{N}$. However, the test requires an operation that changes the ratios between probabilities $p_n$, which can only be done if the multiplicative factors are different for each $n$. Optimally, we would like to have the ratios to be $\frac{f_{n}}{f_{n-1}}=\frac{\sqrt{n+1}}{\sqrt{n}}$. In this case $g(N)$ is a constant, however the normalization constraint and the fact that the sum over $n$ goes to infinity imply that the only possible solution is trivial $g(N)=0$. This problem can be circumvented if the maximal number of particles is bounded. In this case, the optimal operator $a^{\dagger}_{eff}$ is state dependent, i.e., for the state supported on the subspace spanned by $\{|0\rangle,|1\rangle,\dots,|n_{max}\rangle\}$ the corresponding function is  $g(N)=\frac{1}{\sqrt{n_{max}+1}}$, a constant for $n \leq n_{max}$, and $g(N)=\frac{1}{\sqrt{N}}$ for $n>n_{max}$. 
The effective operator $a^{\dagger}_{eff}$ with this function is then the optimal operator to implement particle addition. 

{\it Composite particles of two distinguishable fermions.} We now proceed to situations for which the measure ${\cal M}$ lies between $0$ and $1$. Apart from any experimental imperfections, this may occur if one deals with composite particles, i.e., systems which are composed of an even number of elementary fermions. Here, we discuss the case of a composite particle made of two distinguishable elementary fermions (see Refs. \cite{Law, Wot, Us}), the simplest departure from ideal bosonic behavior. 

The general pure state of two distinguishable fermions can be written as 
\begin{equation}\label{cob}
|\psi\rangle_{AB}=\sum_{k}\sqrt{\lambda_k}a^{\dagger}_k b^{\dagger}_k|0\rangle,
\end{equation}
where $a^{\dagger}_k$ ($ b^{\dagger}_k$) creates a fermion A (B) in mode $k$ and $\lambda_k$ are probabilities. The above state is written in the Schmidt form, i.e., as a sum of tensor products of distinct orthogonal states. The modes $k$ can refer for instance to energy levels of a confining potential, or to the position of the center of mass of A and B. The state can be considered as a single boson state if the operator $c^{\dagger}=\sum_k \sqrt{\lambda_k} a^{\dagger}_k b^{\dagger}_k$ behaves as a proper bosonic creation operator. The corresponding commutation relation reads $[c,c^{\dagger}]=\openone-\Delta$, where $\Delta=\sum_k \lambda_k (a_k^{\dagger}a_k+b_k^{\dagger}b_k)$. The state of $n$ composite bosons is
\begin{equation}
|n\rangle=\chi_n^{-1/2}\frac{c^{\dagger n}}{\sqrt{n!}}|0\rangle, \nonumber
\end{equation}
on which the action of the annihilation operator gives $c|n\rangle=\alpha_n \sqrt{n}|n-1\rangle+|\varepsilon_n\rangle$. The parameters $\chi_n$ and $\alpha_n=\sqrt{\chi_n/\chi_{n-1}}$ are normalization constants and $|\varepsilon_n\rangle$ is a vector orthogonal to $|n-1\rangle$. Moreover,
\begin{equation}\label{eps}
\langle\varepsilon_n|\varepsilon_n\rangle=1-n\frac{\chi_{n}}{\chi_{n-1}}+(n-1)\frac{\chi_{n+1}}{\chi_n}.
\end{equation}
In the ideal scenario $\alpha_n\rightarrow 1$ and $\langle\varepsilon_n|\varepsilon_n\rangle \rightarrow 0$, which happens only if the ratio $\chi_{n\pm 1}/\chi_n \rightarrow 1$. This ratio was related to the parameters $\lambda_k$ and it was shown that it approaches one in the limit of infinite entanglement between A and B \cite{Law,Wot}.

Now, let us examine the action of particle addition for composite particles. One can always write the corresponding Kraus channel as
\begin{equation}
K_0=c^{\dagger}_{eff} =\sum_{n=0}^{n_{max}} g(n+1)\alpha_{n+1}\sqrt{n+1}|n+1\rangle\langle n|,\nonumber 
\end{equation}
where the terms related to $|\varepsilon_n\rangle$ are incorporated in other Kraus operators. The particle subtraction is given by taking hermitian conjugate of the above. The optimal function is again a constant, $g(n+1)=\frac{1}{\alpha_{n_{max}+1}\sqrt{n_{max}+1}}$, for which the ratio 
$\frac{f_{n}}{f_{n-1}}=\frac{\alpha_{n+1}\sqrt{n+1}}{\alpha_{n}\sqrt{n}}$.
The action of addition and subtraction on the optimal initial state $\frac{2}{3}|0\rangle\langle 0| + \frac{1}{3} |1\rangle\langle 1|$ gives
\begin{equation}
{\cal M}=2-\frac{3}{1+2\chi_2^{2}}. \nonumber
\end{equation}
Therefore, one can obtain $\chi_2$ from this test. The purity $P=\text{Tr}\left(\text{Tr}_B(\rho_{AB})\right)^2$, which is a measure of entanglement for pure bipartite states $\rho_{AB}$, is given by $P=1-\chi_2$, and therefore the above test can detect entanglement between the two constituent fermions. For initial states of the form $p_0|0\rangle\langle 0| + (1-p_0)|n\rangle\langle n|$, one can immediately deduce the ratio $\frac{\chi_{n+1}}{\chi_n}$ from ${\cal M}$. Repeating this for all $n$ one can obtain full information about the internal state of the composite particle. We note that a high value of measure ${\cal M}$ for a composite particle implies that the deviation from the bosonic commutation is minimized for states $|n\rangle$ i.e., $\langle n| \Delta | n \rangle$ scales inversely as ${\cal M}$. The measure ${\cal M}$ is thus an indicator of the bosonic quality of composite particles not only with respect to addition and subtraction but also with respect to other bosonic properties such as condensation, bunching etc. 

It is interesting to test our measure on the example of maximally entangled states with $\lambda_k=\frac{1}{d}$, where $d$ is the total number of modes in the system. The corresponding creation operator is
\begin{equation}\label{composite}
c^{\dagger}_{max}=\frac{1}{\sqrt{d}}\sum_{k=1}^d a^{\dagger}_k b^{\dagger}_k. \nonumber
\end{equation}
Note that the consideration of a finite dimension $d$ is natural when one deals with excitations confined in a finite condensed matter region. Moreover, if one deals with particles in free space, due to the finiteness of physical resources such as energy, it is natural to assume finite sums over coefficients $\lambda_k$. In the set of all operators spanned over $d$ fermionic modes, operators $c^{\dagger}_{max}$ are the most bosonic since they correspond to maximally entangled states. Therefore, the bosonic quality of all states in such a set is bounded from above by the bosonic quality of the state defined by $c^{\dagger}_{max}$. For these truncated operators, one is always restricted to states $|n\rangle$ for which $n\leq d$. It is easy to verify that $\langle 0| c^n_{max} c^{\dagger n}_{max}|0\rangle=\frac{n!^2}{d^n}\frac{d!}{n!(d-n)!}$, hence $\chi_n=d!/\left(d^n(d-n)!\right)$ and $\alpha_n^2=(d-n+1)/d$, which implies that the norm in (\ref{eps}) is zero.
The highest quality of bosons obtainable with composite particles is
\begin{equation}
{\cal M}=1-\frac{2d-1}{3d^2-2d+1}.
\end{equation}
In the limit of infinite $d$ (infinite entanglement) the above measure goes to one, which agrees with the results presented in the Refs. \cite{Law, Wot, Us}. 

{\it Composite particles of an even number of fermions.} A general composite particle made of an even number ($2n$) of fermions, such as an atom or an ion, has creation operator given by
\begin{equation}
c^{\dagger}_{multi}=\sum_{\vec{j}}\eta_{\vec{j}} p^{\dagger}_{j_1}\dots p^{\dagger}_{j_k} e^{\dagger}_{j_{k+1}}\dots e^{\dagger}_{j_{l}} n^{\dagger}_{j_{l+1}}\dots n^{\dagger}_{j_{2n}}, \nonumber
\end{equation}
where operators $p^{\dagger}_i$, $e^{\dagger}_j$ and $n^{\dagger}_k$ may be taken to correspond to protons, electrons and neutrons in modes $i$, $j$ and $k$ respectively and $\eta_{\vec{j}}$ is a complex coefficient for a particular configuration $\vec{j}=\{j_1,\dots,j_{2n}\}$. It should be noted that the Kraus channel for addition takes an analogous form to Eq. (\ref{eff}) and the value of the measure ${\cal M}$ is still a good indicator of the bosonic quality of the composite particle. 

As regards the entanglement structure of these composite particles, note that this entanglement is tripartite with respect to the groups of particles of types $p$, $e$ and $n$. Nevertheless, the purity of the reduced density matrix of any single subsystem is a measure of the entanglement between this subsystem and the rest of the particles. For genuine tripartite entanglement of the GHZ type \cite{GHZ}, the measure ${\cal M}$ can still be used to calculate the entanglement in the state. It is an interesting question how much information about other kinds of tripartite entanglement is contained in ${\cal M}$.

{\it Implementation.} So far, we have considered addition and subtraction operations implemented by the optimal Kraus channels. We now discuss possible experimental implementations of these channels. In experiments \cite{Parigi,Zavatta}, photonic addition was implemented by the process of parametric down-conversion (PDC), while photonic subtraction was realized using a beam splitter (BS) with very low reflectivity. In both instances, successful implementation was heralded by a detection event on an auxiliary detector --- for PDC a trigger photon confirmed the addition of the idler photon to the state, whereas for BS the heralding was done by the reflected photon subtracted from the state. The effective addition operator in the process of PDC is
\begin{equation}
a^{\dagger}_{PDC}=\sum_{n=0}^{\infty}\sin(\gamma t \sqrt{n+1})|n+1\rangle\langle n|, \nonumber
\end{equation}
where $\gamma$ is the PDC efficiency and $t$ is the time of interaction between the PDC crystal and the pumping field photon. For small $\gamma$ and $t$, ($t$ is of the order $l/c$, where $l$ is the size of the PDC crystal and $c$ is the speed of light in the crystal) the above process well approximates the optimal bosonic addition channel. Moreover, the effective subtraction operator resulting from a beam splitter of low reflectivity ($r\ll 1$) is of the form
\begin{equation}
a_{BS}=\sum_{n=0}^{\infty}\sqrt{(1-r)r}\sqrt{n+1}|n\rangle\langle n+1|, \nonumber
\end{equation}
which approximates bosonic subtraction very well.

For massive particles, a possible candidate to realize the optimal addition channel is the process of Feshbach resonance, whose effective Hamiltonian is of the form $H_F=\gamma(b^{\dagger}a_1 a_2+b a_1^{\dagger} a_2^{\dagger})$, where $a_1^{\dagger}$ creates an atom of type 1, $a_2^{\dagger}$ an atom of type 2 and $b^{\dagger}$ a two-atom molecule. The optimal subtraction channel for massive particles can be implemented with two potential wells for which the tunneling probability from one to the other is small, effectively mirroring the beam splitter operation. Note that these operations are in principle implementable with current technology with Cs and Rb atoms \cite{N1}.

{\it Discussion.} The measure ${\cal M}$ depends on two crucial features of the addition and subtraction process, namely, the quality of the particle that is to be determined by the test and the quality of the channels used to implement these operations (i.e., how well they approximate the optimal addition/subtraction channel). For the optimal channels, as discussed previously, a good initial state is $p_0=\frac{2}{3}$ and $p_1=\frac{1}{3}$ and the measure takes the form (\ref{measure}). In general, the value of the measure changes with the $f_n$ in the channels that implement the AS operation. Since the test is aimed at determining the quality of particles irrespective of the channel used, we now discuss a way to remove the effects due to imperfect channels. 

For general channels with the fraction $\frac{f_n}{f_{n-1}}\neq \frac{\sqrt{n+1}}{\sqrt{n}}$ it can be easily verified that the measure reads
\begin{equation}
M=p_0-p_0^{AS}=p_0\left(1-\frac{1}{p_0 + (1-p_0)\left(\frac{f_1}{f_0}\right)^4}\right). \nonumber
\end{equation}
The optimal initial state to maximize the value of this measure then assumes the form $p_0=1/\left(1+\left(\frac{f_0}{f_1}\right)^2\right)$ and $p_1 = 1 - p_0$. Therefore, while the optimal initial state differs for different channels the measure ${\cal M}$ is always a valid indicator of bosonic quality irrespective of the addition/subtraction channel used.

{\it Conclusions.} In this paper we proposed a test to determine the bosonic quality of particles by elementary operations of single particle addition and subtraction. We discussed the probabilistic nature of these operations by formulating them within the Kraus operator formalism. By comparing an initial state with one after the process of addition followed by subtraction we formulated a measure that determines the bosonic quality of the particle independently of the quality of channels performing these operations. We also suggested an experimental realization of these channels for composite particles. 

This research is supported by the National Research Foundation and Ministry of Education in Singapore. We acknowledge useful discussions with J. Bandhyopadhyay, R. Fazio, B. Hessmo, T. Paterek and M. Paternostro.


\begin{thebibliography}{99}


\bibitem{Parigi}
V. Parigi, A. Zavatta, M. Kim, M. Bellini 
Science {\bf 317}, 1890 (2007).

\bibitem{Zavatta}
A. Zavatta, V. Parigi, M. Kim, H. Jeong, and M. Bellini
Phys. Rev. Lett. {\bf 103}, 140406 (2009).

\bibitem{NC}
M. A. Nielsen and I. L. Chuang, 
{\it Quantum computation and quantum information}, 
Cambridge University Press, Cambridge (2000).

\bibitem{PKOK}
T. Paterek, P. Kurzynski, D.K.L. Oi and D. Kaszlikowski,
New J. Phys. {\bf 13}, 043027 (2011).

\bibitem{Law}
C. K. Law,
Phys. Rev. A {\bf 71}, 034306 (2005).

\bibitem{Wot}
C. Chudzicki, O. Oke, W. K. Wootters,
Phys. Rev. Lett. {\bf 104}, 070402 (2010).

\bibitem{Us}
R. Ramanathan, P. Kurzynski, T. K. Chuan, M. F. Santos and D. Kaszlikowski,
arXiv:1103.1206 quant-ph (2011).

\bibitem{GHZ}
D. M. Greenberger, M. A. Horne and A. Zeilinger, in: 'Bell's Theorem, Quantum Theory, and Conceptions of the Universe', M. Kafatos (Ed.), Kluwer, Dordrecht, {\bf 69-72} (1989).

\bibitem{N1}
M. Debatin {\it et. al.}, arXiv:1106.0129 physics.atom-ph (2011); 
A. D. Lercher {\it et. al.}, Eur. Phys. J. D (2011), DOI: 10.1140/epjd/e2011-20015-6;
H-C. Nagerl {\it et. al.}, J. Phys.: Conf. Ser. 264 012015 (2011).

\end{thebibliography}
\end{document}